\documentclass[conference]{IEEEtran}
\IEEEoverridecommandlockouts
\usepackage{amsmath,amsfonts,cases}
\usepackage{cite,url}
\usepackage{graphicx}
\newtheorem{proposition}{Proposition}
\newtheorem{definition}{Definition}

\newtheorem{theorem}{Theorem}
\newtheorem{lemma}{Lemma}

\newcommand{\Err}{\mathcal{E}}

\DeclareMathOperator{\EX}{\mathbb{E}}
\DeclareMathOperator{\N}{\mathcal{N}}

\newcommand\numberthis{\addtocounter{equation}{1}\tag{\theequation}}

\begin{document}

    \title{
Secure Semantic Communication over \\
Wiretap Channel

\thanks{This research is supported by the 5/6GIC, Institute for Communication Systems (ICS), University of Surrey and the UK Department for Science, Innovation and Technology under the Future Open Networks Research Challenge project TUDOR (Towards Ubiquitous 3D Open Resilient Network). The views expressed are those of the authors and do not necessarily represent the project.}
}

\author{\IEEEauthorblockN
{
Denis Kozlov,
Mahtab Mirmohseni and Rahim Tafazolli
}
\IEEEauthorblockA
{\\
5/6GIC, Institute for Communication Systems (ICS), University of Surrey, Guildford, United Kingdom
}}

\markboth{}
\IEEEpubid{}

\maketitle
\begin{abstract}
Semantic communication is a new paradigm for information transmission that integrates the essential meaning (semantics) of the message into the communication process.
However, like in classic wireless communications, the open nature of wireless channels poses security risks for semantic communications. In this paper, we characterize information-theoretic limits for the secure transmission of a semantic source over a wiretap channel. Under separate secrecy and distortion constrains for semantics and observed data, we present general inner and outer bounds on the rate-distortion-equivocation region. We also reduce the general region to the case of Gaussian source and Gaussian wiretap channel and provide numerical evaluations. 
\end{abstract}

    \section{Introduction}
Semantic communication is a promising approach for the next generation wireless networks, especially within the realm of 6G technology. In this paradigm, semantic content of messages shapes on the communication process, marking a departure from traditional communication methods where semantic aspects are irrelevant to the transmission problem \cite{qin2022semantic, gunduz2023}. Despite the potential of wireless semantic communication, it still faces substantial security challenges due to the inherent openness of communication channels, leaving the semantic source susceptible to eavesdropping. Furthermore, semantics can carry more sensitive information compared to original data. Thus, the security requirements can be different for the semantics and the original data.

In this paper, we derive the information-theoretic limits governing the secure communication of semantic sources. To achieve this, we model a source composing of two correlated components, the intrinsic (semantic) part and extrinsic (observed) part; building upon previous work that introduced this source model \cite{liu2022indirect, liu2021indirect}. To illustrate this model, an intuitive example is a semantic part represented by a textual description of an image, coupled with an observed part generated by a neural network in response to the text prompt. Our research focuses on a wiretap channel scenario, wherein we consider a passive eavesdropper as the adversary. Within this channel and general semantic source models, we analyse the trade-off between equivocation and distortion for the semantic and observed components separately, particularly in the context of joint source-channel coding (JSCC). Specifically, we derive inner and outer bounds on the rate-distortion-equivocation region, considering two variants of encoder input: only observation; or both observation and semantics. Although, it is reasonable to consider only observation to be encoded, from a practical perspective, the other case allows assessing the gain we get from encoding both.

Wyner's work in 1975 laid the groundwork for secure communication over wiretap channels \cite{wyner1975}, while subsequent advancements generalized this model to broadcast channels with common and confidential messages \cite{csiszar1978}. In \cite{yamamoto1997}, the wiretap model was extended to incorporate JSCC and the one-time pad technique, bringing an important result that a separation principle holds: we can start with a rate-distortion achievable code, followed by one-time pad for a given key-rate, and finish by using a wiretap code. Further extensions to the JSCC model, including scenarios with side information at the decoders, have been explored in \cite{merhav2007, villard2014, villard2011phd}. The lossy compression of semantic sources was covered in \cite{liu2022indirect}. In \cite{stavrou2023jscc}, the authors considered transmission of a semantic source over the noisy channel and derived sufficient conditions for optimal transmission. Secure lossy source coding of semantic sources with secret key is presented in \cite{guo2022}, where secrecy condition is imposed only on semantic part of the data. The model in \cite{guo2022} has similarities with our work. However, taking the step further, we consider the model with \emph{wiretap channel} and secrecy conditions \emph{both} on semantic and observed parts of the source, which allows considering different secure communication setups. For instance, circumstances may require partial or full secrecy for semantics as they carry sensitive information, while secrecy for the observation (carrier of semantics) is non-essential. Our work can be seen as an adaption of model \cite{yamamoto1997} to semantic sources \cite{liu2022indirect}.


Secrecy of semantic sources constitutes trade-off which can be described using information-theoretic approach. To the best of our knowledge, this paper is the first work on secure JSCC of semantic sources over wiretap channel under separate equivocation and distortion conditions for semantic and observed components of the source. This problem is particularly relevant for characterizing security limits of future generation wireless networks such as 6G, where semantic aspects of the source have a greater interest \cite{liu2022indirect}.


\newpage
    \section{Problem statement} \label{section:problem-statement}

\begin{figure}
    \centering
    \includegraphics[width=0.5\textwidth]{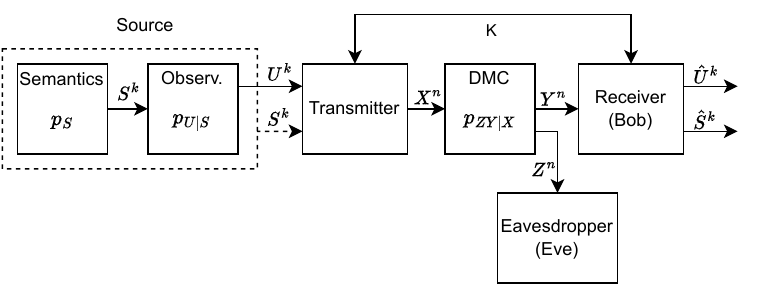}
    \caption{Information-theoretic system model for communication of semantic sources over a wiretap channel with two distortion constrains at the legitimate receiver and two equivocation constrains at the eavesdropper. Two different transmitter input cases are considered: access to both semantic part and observation, and access only to observation.}
    \label{fig:model}
\end{figure}

Consider a model shown in Fig. \ref{fig:model}, where a transmitter wishes to send the semantic and observed data to a legitimate receiver (Bob), subject to some average distortion constrains while keeping them hidden from an eavesdropper (Eve) with some secrecy  (equivocation) constraints. The source consists of an intrinsic (semantic) part and an extrinsic observation part modelled as a sequence of independent and identically distributed (i.i.d) random variables (r.v.s) $S^k$ and $U^k$, respectively, correlated through joint distribution $p_{S,U}(s,u)$ defined on the product alphabet $\mathcal{S} \times \mathcal{U}$.
Main (Bob's) and wiretap (Eve's) channels are modelled by discrete memoryless channels (DMCs) with input $X$ on $\mathcal{X}$ and outputs $Y$ on $\mathcal{Y}$ and $Z$ on $\mathcal{Z}$ given transition probability $p_{Y,Z|X}$.
In this paper, we consider a degraded wiretap channel model, i.e., $p_{Y,Z|X} = p_{Y|X} p_{Z|X}$. The transmitter and receiver have access to a shared key modelled as a random variable $K$ with alphabet $\mathcal{K}$. This key is used to secure a part of the data with a one-time pad technique.

We examine two cases for the transmitter (encoder) input. In the first case, the encoder has observation $u^k$ as an input and has no direct access to a semantic sequence $s^k$, while in the second case, the encoder has access to both semantic $s^k$ and observation $u^k$. The case 1 encoder $f_1 : U^k \times K \rightarrow X^n$ maps the observed sequence and a key to a channel input sequence $X^n$. The case 2 encoder $f_2 : S^k \times U^k \times K \rightarrow X^n$, acts on the observed sequence, semantic sequence, and key.  We define the decoding function as $\hat{f} : Y^n \times K \rightarrow (\hat{S}^k, \hat{U}^k)$
which maps Bob's received signal $Y^n$ and the shared key to estimated semantic and observed parts $\hat{S}^k$ and $\hat{U}^k$ defined on alphabets $\hat{\mathcal{S}}$, $\hat{\mathcal{U}}$. From the decoder function definition, we have:
\begin{equation} \label{eq:key-condition}
    H(\hat{S}^k, \hat{U}^k|Y^n, K) = 0.
\end{equation}
Additionally, we set distortion measures for semantics $d_s : \mathcal{S} \times \mathcal{\hat{S}} \rightarrow [0, \infty)$ and for observation $d_u : \mathcal{U} \times \mathcal{\hat{U}} \rightarrow [0, \infty)$. We follow the similar notation for the block average distortion: $d_s(s^k,\hat{s}^k) = \frac{1}{k} \sum^{k}_{i=1} d_s(s_i,\hat{s}_i)$ and $d_u(u^k,\hat{u}^k) = \frac{1}{k} \sum^{k}_{i=1} d_u(u_i,\hat{u}_i)$.
The goal of this work is to characterize the set $\mathcal{R}$ of all achievable tuples:
\begin{equation*} \label{eq:region-definition}
    \mathcal{R} \doteq \{
    (R, R_k, D_s, D_u, \Delta_s, \Delta_u, \Delta_{su})
        \text{ is achievable}
    \},
\end{equation*}
which will be defined in the following.

\begin{definition} \label{def:code}
Source-channel code $(k,n)$ is defined by a stochastic encoding function $f_i$, where $i \in \{1,2\}$ denotes the encoder input type, and Bob's decoding function $\hat{f}$.
\end{definition}

\begin{definition} \label{def:achievable}
A tuple $(R, R_k, D_s, D_u, \Delta_s, \Delta_u, \Delta_{su}) \in \mathbb{R}^7_+$ is achievable if there exists a source-channel $(k,n)$-code $(f_i, \hat{f})$, for $i \in \{1,2\}$, such that
\begin{align}
\label{eq:condition_for_rate}
n/k &\leq R + \epsilon, \\
\label{eq:condition_for_key_rate}
\frac{1}{k} \log |\mathcal{K}|  &\leq R_k + \epsilon, \\
\label{eq:cond_for_sem_distortion}
\EX d_s(S^k, \hat{S}^k) &\leq D_s + \epsilon, \\
\label{eq:cond_for_src_distortion}
\EX d_u(U^k, \hat{U}^k) &\leq D_u + \epsilon, \\
\label{eq:cond_for_sem_equivocation}
\frac{1}{k} H(S^k|Z^n) &\geq \Delta_s - \epsilon, \\
\label{eq:cond_for_src_equivocation}
\frac{1}{k} H(U^k|Z^n) &\geq \Delta_u - \epsilon, \\
\label{eq:cond_for_joint_equivocation}
\frac{1}{k} H(S^k,U^k|Z^n) &\geq \Delta_{su} - \epsilon
\end{align}
are satisfied for any $\epsilon > 0$.
\end{definition}

Condition (\ref{eq:condition_for_rate}) restricts the number of channel uses per source symbol 
. Equation (\ref{eq:condition_for_key_rate}) sets the condition for the rate of the key which is used to protect data with the one-time pad technique. Average distortions for semantics and observation are separately bounded by conditions (\ref{eq:cond_for_sem_distortion}) and (\ref{eq:cond_for_src_distortion}), respectively. Equivocation for semantic and observed parts as well as joint ones are restricted by conditions (\ref{eq:cond_for_sem_equivocation}), (\ref{eq:cond_for_src_equivocation}) and (\ref{eq:cond_for_joint_equivocation}), respectively.

Next, we provide some relevant notions from rate-distortion theory. Lemma \ref{lemma:classic-rdf} describes rate-distortion function (RDF) for classical source coding setup with single distortion measure, Lemma \ref{lemma:bivar-rdf} refers to RDF for semantic source coding, where encoder having both semantic and observation as an input, and Lemma \ref{lemma:semantic-rdf} corresponds to encoder with only observation as an input.

\begin{lemma}[{\hspace{1sp}\cite[Theorem 3.5]{gamal2011}}]
    The rate-distortion function for discrete memoryless source (DMS) $U$ has the following form 
    \begin{equation*} \label{eq:classic_rate_distortion_src}
    R_u(D_u) = \inf_{\substack{ p_{\hat{U}|U} \\
    \EX d_u(U, \hat{U}) \leq D_u }}
    I(U; \hat{U}).
\end{equation*}
\label{lemma:classic-rdf}
\end{lemma}


\begin{lemma}[{\hspace{1sp}\cite[Theorem 2]{elgamal1982}}]
The rate-distortion function for a bivariate source with separate distortions is given by
\begin{equation*}
    R(D_s, D_u) = \inf_{\substack{ p_{\hat{S}, \hat{U}|S,U} \\
    \EX d_u(U, \hat{U}) \leq D_u \\
    \EX d_s(S, \hat{S}) \leq D_s }}
    I(S, U;\hat{S}, \hat{U}).
\end{equation*}
\label{lemma:bivar-rdf}
\end{lemma}

\begin{lemma}[{\hspace{1sp}\cite[Theorem 1]{liu2022indirect}}]
The semantic rate-distortion function is expressed by
    \begin{equation*}
        R(D_s, D_u) = \inf_{\substack{ p_{\hat{S}, \hat{U}|U} \\
        \EX d_u(U, \hat{U}) \leq D_u \\
        \EX \hat{d}_s(U, \hat{S}) \leq D_s }}
        I(U;\hat{S}, \hat{U}),
    \end{equation*}
    where $\hat{d}_s(U, \hat{S}) = \sum_{s \in S} p_{S|U}(s|U) d_s(s,\hat{S})$ is a modified distortion metric.
\label{lemma:semantic-rdf}
\end{lemma}
    \section{Main Results}
This section presents the general converse and achievable region for a system model defined in Section \ref{section:problem-statement}.
\begin{theorem} \label{theorem:converse}
    (Converse). For both cases of encoder input, given degraded channel $P_{YZ|X}$ and semantic source $P_{SU}$, any achievable tuple $(R, R_k, D_s, D_u, \Delta_s, \Delta_u, \Delta_{su})$ must satisfy:
    \begin{numcases}{}
    R_u(D_u) \leq R I(X;Y),
    \label{eq:rate_region_src} \\
    R_s(D_s) \leq R I(X;Y),
    \label{eq:rate_region_sem} \\
    R(D_s,D_u) \leq R I(X;Y),
    \label{eq:merged_rate_region} \\
    \nonumber
    \Delta_u \leq R_k + R [I(X;Y) - I(X;Z)] \\ \qquad - R_u(D_u) + H(U),
    \label{eq:equivocation_region_src} \\
    \nonumber
    \Delta_s \leq R_k + R [I(X;Y) - I(X;Z)] \\ \qquad - R_s(D_s) + H(S),
    \label{eq:equivocation_region_sem} \\
    \nonumber
    \Delta_{su} \leq R_k + R [I(X;Y) - I(X;Z)] \\ \qquad - R(D_s,D_u) + H(S,U).
    \label{eq:merged_equivocation_region}
    \end{numcases}
    The proof of Theorem \ref{theorem:converse} is provided in Appendix \ref{appendix:converse-proof}.
\end{theorem}

One can see that the equivocation in the converse region for semantics, observation, and both of them comprises three basic terms. The first one, $R_k$, shows how much equivocation we achieve using the one-time pad technique with a key rate $R_k$. The second one is the secrecy capacity $R [I(X;Y) - I(X;Z)]$. And the last one (i.e. $H(U) - R_u(D_u)$ in eq. (\ref{eq:equivocation_region_src})) describes the part of equivocation due to loss in source encoding.

\begin{theorem} \label{theorem:direct}
    (Achievability\footnote{We assume $R_k = 0$ for brevity}). When the transmitter has access to both semantics and observation (case 2 of encoder input), a tuple $(R, D_s, D_u, \Delta_s, \Delta_u, \Delta_{su})$ is achievable if there exist auxiliary r.v.s $A_c,A_p,B_c,B_p,Q_c,Q_p,W_c,X$ with joint distribution $p(a_c,a_p,b_c,b_p,q_c, q_p, w_c, x)$ 
    and functions $\tilde{S} : A^k_c \times A^k_p \rightarrow \hat{S}^k$ and $\tilde{U} : A^k_c \times A^k_p \times B^k_c \times B^k_p \rightarrow \hat{U}^k$ such that the following inequalities hold:
\begin{equation*}
\begin{cases}
    I(S; A_c) < R I(Q_c;Y), \\
    I(S; A_c, A_p) < R I(Y; Q_c, Q_p), \\
    I(U; B_c | S,A_c) < R I(W_c;Y | Q_c), \\
    I(U; B_c | S,A_c) + I(U; B_p | S, A_c, A_p, B_c) < \\
    \qquad < R \left[ I(W_c; Y | Q_c) + I(X;Y | Q_c, Q_p, W_c) \right], \\
    D_s \geq \EX d_S(S, \tilde{S}(A_c, A_p)), \\
    D_u \geq \EX d_U(U, \tilde{U}(A_c, A_p,B_c, B_p)), \\
    \Delta_s \leq H(S|A_c,A_p) + R \big[ H(Z|X)-H(Z|Q_c) ) \\
    \qquad + I(Q_p;Y|Q_c) \big], \\
    \Delta_u \leq H(U) - I(A_c,A_p;S)-I(B_c;U|A_c) \\ 
    \qquad - I(B_p;U|S,A_c,A_p,B_c) + R \big[ H(Z|X) \\ 
    \qquad - H(Z|Q_c,W_c) + I(Q_p;Y|Q_c) \\
    \qquad + I(X;Y|Q_c,Q_p,W_c) \big], \\
    \Delta_{su} \leq H(S,U) - I(S;A_c,A_p) - I(U;B_p|S,A_c,A_p,B_c) \\
    \qquad - I(U;B_c|S,A_c) + R \big[ H(Z|X) - H(Z|Q_c,W_c) \\
    \qquad + I(Q_p;Y|Q_c) + I(X;Y|Q_c,Q_p,W_c) \big].
\end{cases}
\end{equation*}

\textbf{Achievability proof outline.} 
Consider a codebook with source, channel, and wiretap codes. The wiretap code is embedded in the channel encoder and introduces additional random noise to protect private parts of the data from the eavesdropper. We introduce source encoder auxiliary r.v.s $A_c$, $A_p$, $B_c$, $B_p$, and channel encoder auxiliary r.v.s $Q_c$, $Q_p$, $W_c$ for codebook generation. The r.v.s $A_c$ and $A_p$ reflect the distribution of i.i.d. codewords (denoted by $a^k_c$ and $a^k_p$) for the common and private parts of semantics, while $B_c$ and $B_p$ are used for the common and private parts of the observation, for codewords denoted as $b^k_c$ and $b^k_p$, correspondingly. Then we use a technique similar to superposition coding: for each sequence $a^k_c$ we generate $a^k_p$ and $b^k_c$, then for each $a^k_c$, $a^k_p$, $b^k_c$ we generate $b^k_p$. All of these sequences are selected from typical sets. The channel encoder codebook has a layered structure related to the source encoder. That is, we generate sequence $q^n_c$ from $Q_c$ distribution, for each $q^n_c$ we generate $q^n_p$ and $w^n_c$ using $Q_p$ and $W_c$, given $q^n_c$, $q^n_p$, $w^n_c$ we generate $x^n$ according to $X$. Sequences $q^n_p$ and $x^n$ additionally covered with noise. To encode data, we choose source encoder sequences that are jointly typical with encoder input. Then we obtain the channel code sequences and transmit $x^n$. The decoding is based on joint typically with channel output $y^n$.

For this codebook and encoding/decoding procedure, we show that the probability of encoding/decoding errors goes to zero under some rate conditions with $k \rightarrow \infty$. Finally, we bound the average distortion and equivocations. For the complete proof of achievability see Appendix \ref{appendix:direct-proof}.
\end{theorem}
    \section{Gaussian Case}
In this section, we consider the transmission of a bivariate Gaussian source with quadratic distortion measure over the Gaussian wiretap channel with respective noise powers $P_{N_1}$ and $P_{N_2}$ for Bob and Eve, respectively. The channel input has a power limit $P$.

\subsection{System model} \label{sub:gauss_model}
Let source be distributed according to normal distribution $(S, U) \sim \mathcal{N}(0,K)$ with covariance matrix
\begin{equation*} 
K = \begin{pmatrix}
    P_s     & P_{su} \\
    P_{su} & P_u
\end{pmatrix},
\end{equation*}
and correlation coefficient $\rho = \frac{P_{su}}{\sqrt{P_s P_u}}$. We set distortion measures $d_s(x,y) = d_u(x, y) = (x - y)^2$,
and we model the channel as
\begin{align*}
    &\EX(X^2) \leq P, \\
    &Y = X + N_1, \quad N_1 \sim \N(0,P_{N_1}), \\
    &Z = Y + N_2, \quad N_2 \sim \N(0,P_{N_2}),
\end{align*}
where $P$ is the power constraint for channel input, $P_{N_1}$ is the noise power for the main channel, and $P_{N_2}$ is the noise power for the eavesdropper channel, we define $P_N = P_{N_1} + P_{N_2}$.

\subsection{Rate-distortion-equivocation region}
The following Gaussian converse and achievable regions are obtained from Theorem \ref{theorem:converse} and Theorem \ref{theorem:direct}, respectively.
\begin{proposition}[Gaussian Converse]
    \label{prop:gauss_outer}
    For the Gaussian system model defined in \ref{sub:gauss_model} and in case when encoder has access only to $u^k$ (case 1 of encoder input), to fulfil conditions (\ref{eq:condition_for_rate})-(\ref{eq:cond_for_joint_equivocation}) any code must satisfy:
\begin{equation*}
    \begin{cases}
    D_s > \eta \\
    \max \left\{
    \frac{1}{2} \log^+ \frac{P_u}{D_u},
    \frac{1}{2} \log^+ \frac{P_{su}^2}{P_u (D_s - \eta)}
    \right\} \leq \frac{1}{2} R \log \left( 1 + \frac{P}{P_{N_1}} \right) \\
    \nonumber
    \Delta_u \leq R_k + R C_s + \frac{1}{2} \log 2 \pi e D_u \\
    \nonumber
    \Delta_s \leq R_k + R C_s - \frac{1}{2} \log^+ \frac{P_{su}^2}{ 2 \pi e P_s P_u (D_s - \eta) } \\
    \nonumber
    \Delta_{su} \leq R_k + R C_s - \max \left\{
    \frac{1}{2} \log^+ \frac{P_u}{D_u},
    \frac{1}{2} \log^+ \frac{P_{su}^2}{P_u (D_s - \eta)}
    \right\} \\ \qquad + \frac{1}{2} \log (2 \pi e)^2 |K|,
    \end{cases}
\end{equation*}
where $\eta = P_s - \frac{P_{su}^2}{P_u}$, and $C_s = \frac{1}{2} \log \frac{P_N(P + P_{N_1})}{P_{N_1} (P + P_N)}$ is the secrecy channel capacity. We use the closed-form solution for the rate-distortion function obtained in reference {\cite[Proposition 3]{liu2021}}.
\end{proposition}

\begin{proposition}[Gaussian Achievability]
\label{prop:gauss_inner}
When the encoder has access to both $u^k$ and $s^k$ (case 2 of encoder input), for the Gaussian source and channel, the tuple $(R, D_s, D_u, \Delta_s, \Delta_u, \Delta_{su})$ is achievable if:
\begin{equation*}
\begin{cases}
\left(
1 + \frac{ \alpha_1^2 P_s }
         { P_{\tilde{A}_p} }
\right)
\leq
\left(
1 + \frac{ P_{Q_c} + P_{\tilde{Q}_p} }
{ P_{\tilde{W}_c} + P_{\tilde{X}} + P_{N_1} } 
\right)^R, \\
\left(
1 + \frac{ \alpha_2^2 |K| }
         { P_S P_{\tilde{B}_p} }
\right)
\leq
\left(
1 + \frac{ P_{\tilde{W}_c} }
         { P_{\tilde{X}} + P_{\tilde{Q}_p} + P_{N_1} }
\right)^R
\left(
1 + \frac{ P_{\tilde{X}} }
         { P_{N_1} }
\right)^R, \\
D_s \geq 
P_S - \frac{ \alpha_1^2 P_S^2 }
           { \alpha_1^2 P_S + P_{\tilde{A}_p} }, \\
D_u \geq 
P_U - \frac{ \alpha_2^2 P_U^2 }
           { \alpha_2^2 P_U + P_{\tilde{B}_p} }, \\
%
\Delta_s \leq 
\frac{1}{2} \log \left(
2 \pi e \frac{ P_S P_{\tilde{A}_p}}
                     {\alpha_1^2 P_S + P_{\tilde{A}_p}} 
\right)  \\
\qquad + \frac{R}{2} \log \left(
\frac{P_{N}(P - P_{Q_c})}
     {(P_{\tilde{W}_c} + P_{\tilde{X}} + P_{N_1})(P - P_{Q_c} + P_N)}
\right),
\\
\Delta_u \leq
\frac{1}{2} \log
\left(
\frac{ 2 \pi e P_S P_U P_{\tilde{A}_p} P_{\tilde{B}_p} }
{ (\alpha_1^2 P_S + P_{\tilde{A}_p}) (\alpha_2^2 |K| + P_S P_{\tilde{B}_p}) }
\right) \\
\qquad + \frac{R}{2} \log
\left(
\frac{ P_N (P_{\tilde{X}} + P_{\tilde{N}_1}) ( P_{\tilde{Q}_p} + P_{\tilde{W}_c} + P_{\tilde{X}} + P_{\tilde{N}_1}) }
     { P_{\tilde{N}_1} ( P_{\tilde{W}_c} + P_{\tilde{X}} + P_{\tilde{N}_1} ) ( P_{\tilde{Q}_p} + P_{\tilde{X}} + P_N ) }
\right)
, \\
\Delta_{su} \leq \frac{1}{2} \log
\left( 
\frac{ (2\pi e)^2 |K| P_S P_{\tilde{A}_p} P_{\tilde{B}_p} }
     { (\alpha_1^2 P_S + P_{\tilde{A}_p}) (\alpha_2^2 |K| + P_S P_{\tilde{B}_p}) }
\right) \\
\quad + \frac{R}{2} \log 
\left(
\frac{ P_N (P_{\tilde{X}} + P_{\tilde{N}_1}) ( P_{\tilde{Q}_p} + P_{\tilde{W}_c} + P_{\tilde{X}} + P_{\tilde{N}_1}) }
     { P_{\tilde{N}_1} ( P_{\tilde{W}_c} + P_{\tilde{X}} + P_{\tilde{N}_1} ) ( P_{\tilde{Q}_p} + P_{\tilde{X}} + P_N ) }
\right).
\end{cases}
\end{equation*}
\end{proposition}

The Gaussian inner bound is derived from Theorem \ref{theorem:direct} by choosing the following auxiliary r.v.s.
The source encoder variables: $A_c = \emptyset, \quad B_c = \emptyset$, $A_p = \alpha_1 S + \tilde{A}_p$, where $\tilde{A}_p \sim \N(0,P_{\tilde{A}_p})$ and $B_p = \alpha_2 U + \tilde{B}_p + \gamma S$, given $\tilde{B}_p \sim \N(0,P_{\tilde{B}_p})$. 

The channel encoder variables are as follows,
\begin{align*}
    &Q_c \sim \N(0,P_{Qc}), \\
    &Q_p = Q_c + \tilde{Q_p},
    \quad \tilde{Q}_p \sim \N(0,P_{\tilde{Q}_p}), \\
    &W_c = Q_c + \tilde{W_c},
    \quad \tilde{W}_c \sim \N(0,P_{\tilde{W}_c}), \\
    &X = Q_c + \tilde{W_c} + \tilde{Q_p} + \tilde{X},
    \quad \tilde{X} \sim \N(0,P_{\tilde{X}}),
\end{align*}
where $P = P_{\tilde{X}} + P_{Qc} + P_{\tilde{W}_c} + P_{\tilde{Q}_p}$.

The function $\tilde{S}$ is defined as minimum mean square error (MMSE) estimator of $S$ from $A_c$ and $A_p$, and $\tilde{U}$ is the MMSE estimation of $U$ from $B_c, B_p$.

\subsection{Numerical evaluation}

\begin{figure}
	\centering
\includegraphics[width=0.47\textwidth]{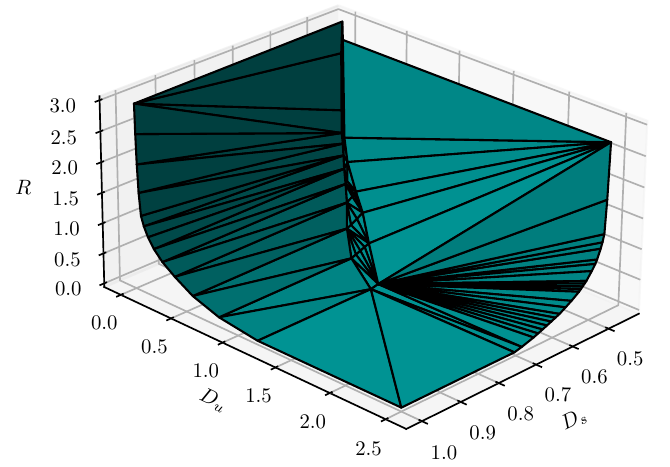}
    \caption{Converse bound for Gaussian system model with full semantic secrecy. The encoder has access only to the observed part.}
    \label{fig:gauss_outer}
\end{figure}

Fig. \ref{fig:gauss_outer} shows the outer bound obtained numerically from Proposition \ref{prop:gauss_outer} for the full semantic security case, where $\Delta_s = H(S)$, $\Delta_u = 0$ and $\Delta_{su}=H(S)$. The other parameters used for Fig. \ref{fig:gauss_outer} are: $R_k = 0$ (no secret key), $P_s = 0.7$, $P_u = 1.0$, $P_{su} = 0.5$, $P = 1$, $P_{N_1} = 0.10$, $P_{N_2} = 0.15$.

\begin{figure}
	\centering
\includegraphics[width=0.47\textwidth]{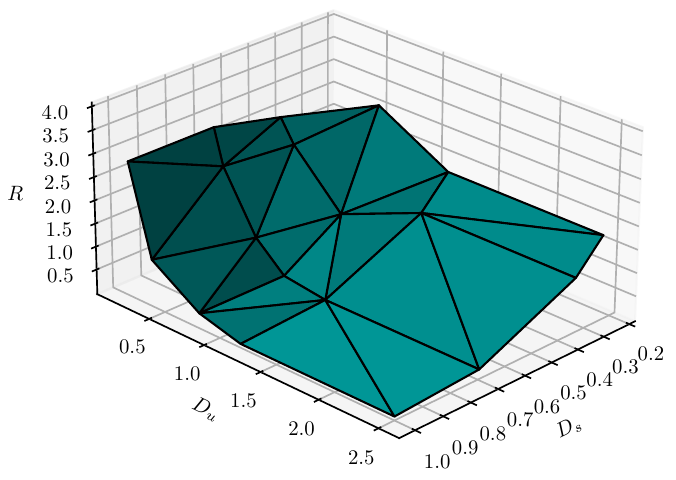}
    \caption{Achievable bound for Gaussian system model with full semantic secrecy. The encoder has access to both semantic and observed part.}
    \label{fig:gauss_inner}
\end{figure}

For the same parameter set, Fig. \ref{fig:gauss_inner} presents an achievable bound, which derived only for the second case of the encoder input, so we cannot compare it with the converse bound.
    \section{Conclusion}
This paper presented an information-theoretic framework for secure transmission of semantic sources over a wiretap channel to examine the theoretically feasible trade-off between distortion and secrecy for semantic sources. We established bounds on the general inner and outer rate-distortion-equivocation region and reduced the converse and achievable bounds to the Gaussian models of source and channel to provide numerical results.

\newpage
    
    \begin{appendices}
        \section{Proof of Theorem \ref{theorem:converse}}
\label{appendix:converse-proof}

The following set of inequalities will be used in our proof,
\begin{align}
    \label{eq:ch-ineq-chain}
	&I(X^n; Y^n) \leq \sum_{i=1}^{n} I(X_i; Y_i) \leq n I(X;Y), \\
    \label{eq:src-ineq-chain}
	&\frac{1}{k} I(U^k; \hat{U}^k) \geq \frac{1}{k} \sum_{i=1}^{k} I(U_i; \hat{U}_i) \geq R_u(D_u + \epsilon), \\
    \label{eq:sem-ineq-chain}
	&\frac{1}{k} I(S^k; \hat{S}^k) \geq \frac{1}{k} \sum_{i=1}^{k} I(S_i; \hat{S}_i) \geq R_s(D_s + \epsilon), \\
    \label{eq:sem-dpi}
    &\sum_{i=1}^{k} I(S_i; \hat{S}_i)
    \leq \sum_{i=1}^{n} I(X_i; Y_i), \\
    \label{eq:src-dpi}
    &\sum_{i=1}^{k} I(U_i; \hat{U}_i)
    \leq \sum_{i=1}^{n} I(X_i; Y_i),
\end{align}
where $\epsilon > 0$ \cite{gamal2011}. Proof of (\ref{eq:sem-dpi}) and (\ref{eq:src-dpi}) is provided in Appendix \ref{appendix:lemmas}. First, we prove the inequality in (\ref{eq:rate_region_src}) as
\begin{align*}
    \nonumber
	R_u(D_u + \epsilon)
    &\leq^{ (\ref{eq:src-ineq-chain}) }
    \frac{1}{k} \sum_{i=1}^k I(U_i;\hat{U}_i)
    \leq^{ (\ref{eq:src-dpi}) }
    \frac{1}{k} \sum_{i=1}^n I(X_i;Y_i) \\
    &\leq^{ (\ref{eq:ch-ineq-chain}) }
    \frac{n}{k} I(X;Y)
    \leq^{ (\ref{eq:condition_for_rate}) }
    (R + \epsilon) I(X;Y).
\end{align*}
To obtain (\ref{eq:rate_region_sem}), we follow the same steps as in the proof of (\ref{eq:rate_region_src}) using (\ref{eq:sem-ineq-chain}) instead of  (\ref{eq:src-ineq-chain}), and (\ref{eq:sem-dpi}) instead of (\ref{eq:src-dpi}).

Next, the proof of (\ref{eq:equivocation_region_src}) is as follows.
\begin{align*}
    &k(R_k + \epsilon)
    \geq^{(\ref{eq:condition_for_key_rate})}
    \log |\mathcal{K}| \geq H(K) \geq H(K|Y^n) \\
    &\geq H(K|Y^n) - H(K|Y^n \hat{U}^k) = H(\hat{U}^k|Y^n) - H(\hat{U}^k|Y^n K) \\
    &=^{(\ref{eq:key-condition})} H(\hat{U}^k|Y^n) \geq^{(\ref{eq:cond_for_src_equivocation})}
    H(\hat{U}^k|Y^n) - H(U^k|Z^n) + k(\Delta_u - \epsilon) \\
    &= H(\hat{U}^k Y^n) - H(Y^n) - H(U^k Z^n) + H(Z^n) + k(\Delta_u - \epsilon) \\
    &= I(U^k; Z^n) - I(\hat{U}^k;Y^n) + H(\hat{U}^k|U^k) - H(U^k) \\
    &\quad + I(U^k;\hat{U}^k) + k(\Delta_u - \epsilon).
    \numberthis\label{eq:inter-result}
\end{align*}
The first three terms in (\ref{eq:inter-result}) can be rewritten as follows:
\begin{align*}
    &I(U^k; Z^n) - I(\hat{U}^k; Y^n) + H(\hat{U}^k|U^k) \\
    &=I(U^k; Z^n) -  H(Y^n) + H(Y^n|\hat{U}^k) + H(\hat{U}^k|U^k) \\
    &\geq I(U^k; Z^n) -  H(Y^n) + H(Y^n|\hat{U}^k U^k) + H(\hat{U}^k|U^k) \\
    &= I(U^k; Z^n) -  H(Y^n) + H(Y^n, \hat{U}^k|U^k) \\
    &\geq I(U^k;Z^n) - I(U^k;Y^n) \geq^{(a)} n [I(X;Z) - I(X;Y)], \numberthis\label{eq:csiszar-ineq}
\end{align*}
where (a) is due to Lemma \ref{lemma:csiszar}.
Substituting (\ref{eq:csiszar-ineq}) to (\ref{eq:inter-result}) we obtain:
\begin{align*}
    \nonumber
    &k(R_k + \epsilon)
    \geq^{(\ref{eq:inter-result}, \ref{eq:src-ineq-chain})}
    n [I(X;Z) - I(X;Y)] - k H(U) \\
    &\quad + k R_u(D_u + \epsilon) + k(\Delta_u - \epsilon) \\
    &\geq^{(\ref{eq:condition_for_rate})} 
    (R + \epsilon) [I(X;Z) - I(X;Y)] - H(U) \\
    &\quad + R_u(D_u + \epsilon) + \Delta_u.
\end{align*}

To prove (\ref{eq:equivocation_region_sem}), we skip some steps due to its similarity with steps in proof of (\ref{eq:equivocation_region_src}) (we use $S^k$ instead of $U^k$).
\begin{align*}
    &k(R_k + \epsilon)
    \geq ...\geq H(\hat{S}^k|Y^n) - H(\hat{S}^k|Y^n K) \\
    &\geq^{(\ref{eq:cond_for_sem_equivocation})}
    H(\hat{S}^k|Y^n) - [H(S^k|Z^n) - k(\Delta_s - \epsilon)] \\
    &= I(S^k; Z^n) - I(\hat{S}^k;Y^n) - H(S^k) + I(S^k;\hat{S}^k) + H(\hat{S}^k|S^k) \\
    &\quad + k(\Delta_s - \epsilon) \\
    &\geq n[I(X;Z) - I(X;Y)] - k H(S) + I(S^k; \hat{S}^k) + k(\Delta_s - \epsilon) \\
    &\geq^{(\ref{eq:sem-ineq-chain})}
    n [I(X;Z) - I(X;Y)] - k H(S) + k R_s(D_s + \epsilon) \\
    &\quad + k(\Delta_s - \epsilon).
\end{align*}

Now we proceed with the proof of (\ref{eq:merged_rate_region}). Our proof relies on the following inequality:
\begin{align} \label{eq:inequality_indirect_rd}
    \frac{1}{k} I(S^k, U^k;\hat{S}^k, \hat{U}^k) \geq \frac{1}{k} \sum_{i=1}^{k} I(S_i, U_i;\hat{S}_i, \hat{U}_i)
    \geq R(D_s, D_u).
\end{align}

Thus, we have:
\begin{align*}
	&R(D_s + \epsilon, D_u + \epsilon)
    \leq^{(\ref{eq:inequality_indirect_rd})}
    \frac{1}{k} \sum_{i=1}^k I(S_i, U_i;\hat{S}_i,\hat{U}_i) \\
    &\leq^{(a)} \frac{1}{k} \sum_{i=1}^n I(X_i;Y_i)
    \leq^{(\ref{eq:ch-ineq-chain})}
    \frac{n}{k} I(X;Y) \leq^{ (\ref{eq:condition_for_rate}) }
    (R + \epsilon) I(X;Y),
\end{align*}
where (a) is due to the data processing inequality.

Finally, the proof of (\ref{eq:merged_equivocation_region}) is,
\begin{align*}
    &k(R_k + \epsilon) \geq ...
    \geq H(K|Y^n) - H(K|Y^n, \hat{S}^k, \hat{U}^k) \\
    %
    &= H(\hat{S}^k, \hat{U}^k|Y^n) - H(\hat{S}^k, \hat{U}^k|Y^n K)
    =^{(\ref{eq:key-condition})}
    H(\hat{S}^k, \hat{U}^k|Y^n) \\
    &\geq^{(\ref{eq:cond_for_joint_equivocation})}
    H(\hat{S}^k,\hat{U}^k|Y^n) - [H(S^k,U^k|Z^n) - k(\Delta_{su} - \epsilon)] \\
    &= H(\hat{S}^k,\hat{U}^k,Y^n) - H(Y^n) - H(S^k,U^k,Z^n) + H(Z^n) \\
    &\quad + k(\Delta_{su} - \epsilon) \\
    &= H(Y^n|\hat{S}^k,\hat{U}^k) + H(\hat{S}^k,\hat{U}^k) - H(Y^n) - H(Z^n|S^k,U^k) \\
    &\quad - H(S^k,U^k) + H(Z^n) + k(\Delta_{su} - \epsilon) \\
    &= I(S^k,U^k; Z^n) - I(\hat{S}^k,\hat{U}^k;Y^n) + H(\hat{S}^k,\hat{U}^k|S^k,U^k) \\
    &\quad - H(S^k,U^k) + I(S^k,U^k;\hat{S}^k,\hat{U}^k) + k(\Delta_{su} - \epsilon).
    \numberthis\label{eq:deltaSU-inter-result}
\end{align*}
The first three terms in (\ref{eq:deltaSU-inter-result}) are,
\begin{align}
\nonumber
    &I(S^k,U^k; Z^n) - I(\hat{S}^k,\hat{U}^k;Y^n) + H(\hat{S}^k,\hat{U}^k|S^k,U^k) \\
    &\geq n[I(X;Z) - I(X;Y)].
    \label{eq:secrecy-capacity-dpi}
\end{align}
Substituting (\ref{eq:secrecy-capacity-dpi}) in (\ref{eq:deltaSU-inter-result}), we have:
\begin{align*}
    &k(R_k + \epsilon) \geq^{(\ref{eq:inequality_indirect_rd})}
    n[I(X;Z) - I(X;Y)] - kH(S,U) \\
    &\qquad + k R(D_s,D_u) + k(\Delta_{su} - \epsilon),
\end{align*}
\begin{align*}
    &R_k + 2\epsilon \geq
    (R+\epsilon)[I(X;Z) - I(X;Y)] - H(S,U) \\
    &\qquad + R(D_s,D_u) + \Delta_{su}.
\end{align*}
Letting $\epsilon \rightarrow 0$ completes the converse proof.
        \section{Proofs of Lemma \ref{lemma:secrecy-capacity-dpi} and Lemma \ref{lemma:csiszar}}
\label{appendix:lemmas}

\begin{lemma} \label{lemma:secrecy-capacity-dpi}
The following two inequalities are valid for the system model described in Section \ref{section:problem-statement}.

\begin{align*}
    \sum_{i=1}^{k} I(S_i; \hat{S}_i)
    \leq \sum_{i=1}^{n} I(X_i; Y_i), \\
    \sum_{i=1}^{k} I(U_i; \hat{U}_i)
    \leq \sum_{i=1}^{n} I(X_i; Y_i).
\end{align*}
\end{lemma}
\textbf{Proof.}
From the system model in Fig. \ref{fig:model}, we have the following Markov Chains

\begin{align}
\label{eq:sem-markov-chain}
   S^k \rightarrow X^n \rightarrow Y^n \rightarrow \hat{S}^k, \\
\label{eq:src-markov-chain}
   U^k \rightarrow X^n \rightarrow Y^n \rightarrow \hat{U}^k. 
\end{align}
Due to memoryless property of the channel we have \cite{gamal2011}
\begin{equation}
\label{eq:channel-mi-inequality}
    I(X^n;Y^n) \leq \sum_{i=1}^{n} I(X_i;Y_i).
\end{equation}
Due to data processing inequality (DPI) from (\ref{eq:sem-markov-chain}) and (\ref{eq:src-markov-chain}) we have
\begin{align}
    I(S^k;\hat{S}^k) \leq I(X^n;Y^n), \\
    I(U^k;\hat{U}^k) \leq I(X^n;Y^n).
\end{align}
In addition,
\begin{align*}
    &I(S^k;\hat{S}^k) = H(S^k) - H(S^k|\hat{S}^k) \\
    =^{(a)} &\sum_{i=1}^{k} H(S_i) -  \sum_{i=1}^{k} H(S_i|\hat{S}^k, S_1,...,S_{i-1}) \\
    \geq &\sum_{i=1}^{k} H(S_i) - \sum_{i=1}^{k} H(S_i|\hat{S}_i) \\
    = &\sum_{i=1}^{k} \left[ H(S_i) - H(S_i|\hat{S}_i) \right] = \sum_{i=1}^{k} I(S_i;\hat{S}_i)
\end{align*}
where (a) because $S^k$ is i.i.d, and due to chain rule for conditional entropy.
Following the same reasoning for $I(U^k;\hat{U}^k)$ we obtain
\begin{align*}
    &I(U^k;\hat{U}^k) = H(U^k) - H(U^k|\hat{U}^k) \\
    =^{(a)} &\sum_{i=1}^{k} H(U_i) -  \sum_{i=1}^{k} H(U_i|\hat{U}^k, U_1,...,U_{i-1}) \\
    \geq &\sum_{i=1}^{k} H(U_i) - \sum_{i=1}^{k} H(U_i|\hat{U}_i) \\
    = &\sum_{i=1}^{k} \left[ H(U_i) - H(U_i|\hat{U}_i) \right] = \sum_{i=1}^{k} I(U_i;\hat{U}_i),
\end{align*}
where (a) because $U^k$ is i.i.d. So we have
\begin{align}
\label{eq:tensor-sem}
    \sum_{i=1}^{k} I(S_i; \hat{S}_i) \leq I(S^k;\hat{S}^k), \\
\label{eq:tensor-src}
    \sum_{i=1}^{k} I(U_i; \hat{U}_i) \leq I(U^k;\hat{U}^k).
\end{align}
The end of the proof as follows
\begin{align*}
    \sum_{i=1}^{k} I(S_i; \hat{S}_i) \leq^{(\ref{eq:tensor-sem})} I(S^k;\hat{S}^k) \\ \leq^{(\ref{eq:sem-dpi})} I(X^n;Y^n) \leq^{(\ref{eq:channel-mi-inequality})} \sum_{i=1}^{n} I(X_i;Y_i), \\
    \sum_{i=1}^{k} I(U_i; \hat{U}_i) \leq^{(\ref{eq:tensor-src})} I(U^k;\hat{U}^k)  \\
    \leq^{(\ref{eq:src-dpi})} I(X^n;Y^n) \leq^{(\ref{eq:channel-mi-inequality})} \sum_{i=1}^{n} I(X_i;Y_i).
\end{align*}

\begin{lemma} \label{lemma:csiszar}
For $U^k \rightarrow X^n \rightarrow Y^n \rightarrow Z^n$ Markov chain, the following holds \cite{csiszar1978}:
\begin{equation}
    I(U^k;Z^n) - I(U^k;Y^n) \geq n [I(X;Z) - I(X;Y)].
\end{equation}
\end{lemma}
\textbf{Proof.} The following is valid due to chain rule for mutual information:
\begin{align*}
    I(U^k; Y^n) = \sum_{i=1}^{n} I(U^k; Y_i|Y^{i-1})
\end{align*}
The term under summation as follows,
\begin{align*}
    &I(U^k; Y_i|Y^{i-1}) 
    = I(U^k, \tilde{Z}^{i+1}; Y_i|Y^{i-1})  - I(\tilde{Z}^{i+1}; Y_i|Y^{i-1}, U^k) \\
    &= I(U^k; Y_i | Y^{i-1}, \tilde{Z}^{i+1}) + I(\tilde{Z}^{i+1}; Y_i | Y^{i-1}) \\
    &\qquad - I(\tilde{Z}^{i+1}; Y_i | U^k, Y^{i-1}),
\end{align*}
where $\tilde{Z}^{i} = (Z_{i}, Z_{i+1}, ..., Z_{n})$.
Following the same steps for $I(U^k; Z^n)$, we have:
\begin{align*}
    I(U^k; Y^n) = \sum_{i=1}^{n} A_i + \sum_{i=1}^{n} B_i - \sum_{i=1}^{n} C_i, \\
    I(U^k; Z^n) = \sum_{i=1}^{n} D_i + \sum_{i=1}^{n} F_i - \sum_{i=1}^{n} G_i,
\end{align*}
where:
\begin{align*}
    A_i &= I(U^k; Y_i | Y^{i-1}, \tilde{Z}^{i+1}), \\
    B_i &= I(\tilde{Z}^{i+1}; Y_i | Y^{i-1}), \\
    C_i &= I(\tilde{Z}^{i+1}; Y_i | U^k, Y^{i-1}), \\
    D_i &= I(U^k; Z_i | Y^{i-1}, \tilde{Z}^{i+1}), \\
    F_i &= I(Y^{i-1}; Z_i | \tilde{Z}^{i+1}), \\
    G_i &= I(Y^{i-1}; Z_i | \tilde{Z}^{i+1} U^k). \\
\end{align*}
Due to Csiszár Sum Identity \cite{gamal2011} we have:
\begin{align*}
    \sum_{i=1}^{n} B_i = \sum_{i=1}^{n} F_i \text{  and  }
    \sum_{i=1}^{n} C_i = \sum_{i=1}^{n} G_i.
\end{align*}
Then:
\begin{align}
    \nonumber
    &I(U^k; Z^n) - I(U^k; Y^n) \\
    \nonumber
    &= \sum_{i=1}^{n} I(U^k; Z_i | Y^{i-1}, \tilde{Z}^{i+1}) - \sum_{i=1}^{n} I(U^k; Y_i | Y^{i-1}, \tilde{Z}^{i+1}) \\
    &= nI(U^k;Z | V) - nI(U^k;Y | V),
    \label{eq:diff_of_MI}
\end{align}
where $J$ is uniform r.v. with alphabet $\mathcal{J}=\{1,...,n\}$, $Y = Y_J$, $Z = Z_J$, and $V = (Y^{i-1}, \tilde{Z}^{i+1}, J)$.

Due to $V \rightarrow W \rightarrow X \rightarrow Y \rightarrow Z$ Markov chain, where $W=(V,U^k)$, we can rewrite (\ref{eq:diff_of_MI}) as,
\begin{align*}
    &nI(W;Z | V) - nI(W;Y | V) = n [I(W;Z) - I(V;Z) \\ &\qquad - I(W;Y) + I(V;Y)] \\
    &= n [I(X;Z) - I(X;Z|W) - I(V;Z) \\
    & \qquad - I(X;Y) + I(X;Y|W) + I(V;Y)] \\ 
    &\geq n [I(X;Z) - I(X;Y)] 
\end{align*}
The proof is complete.
        \section{Proof of Theorem \ref{theorem:direct}}
\label{appendix:direct-proof}

Now we consider achievability proof for case 2 of encoder input (encoder has access to both $u^k$ and $s^k$).

\textbf{Source codebook}. We introduce 4 r.v.s $A_c$, $A_p$ and $B_c$, $B_p$ defined on alphabets $\mathcal{A}_c$, $\mathcal{A}_p$ and $\mathcal{B}_c$, $\mathcal{B}_p$. Random variables $A_c$ and $B_c$ correspond to a codebook distribution of a common message part for semantics and source. While $A_p$ and $B_p$ represent private part of semantics and source. Let $R_{ac}$, $R_{ap}$, $R_{bc}$, $R_{bp}$ be positive rates.

To construct a source codebook, we start by randomly and independently picking $2^{k R_{ac}}$ typical $\mathcal{T}^k_{\delta}(A_c)$ sequences from $A_c$ distribution. We call such sequence $a_c^k(s_{ac})$, where $s_{ac} \in [1,2,...,2^{kR_{ac}}]$

For each sequence $a_c^k(s_{ac})$ we pick $2^{kR_{ap}}$ typical $\mathcal{T}^k_{\delta}(A_p | a_c^k(s_{ac}))$ sequences from $A_p$ distribution and name it $a^k_p(s_{ac},s_{ap})$, $s_{ap} \in [1,2,...,2^{kR_{ap}}]$.

For each $a_c^k(s_{ac})$ we pick $2^{kR_{bc}}$ sequences $b^k_c(s_{ac},s_{bc}) \in \mathcal{T}^k_{\delta}(B_c | a_c^k(s_{ac}))$, $s_{bc} \in [1,2,...,2^{kR_{bc}}]$.

We finish codebook by picking for each previous sequences,
$2^{kR_{bp}}$ sequences $b^k_p(s_{ac}, s_{ap}, s_{bc}, s_{bp}) \in \mathcal{T}^k_{\delta}(B_p | a_c^k(s_{ac}), a^k_p(s_{ac},s_{ap}), b^k_c(s_{ac},s_{bc}))$, $s_{bp} \in [1,2,...,2^{kR_{bp}}]$.

This codebook is revealed to Bob and Eve.

\textbf{Channel codebook}. Let $Q_c$, $Q_p$ and $W_c$ be r.v. for channel codebook generation defined on $\mathcal{Q}_c$, $\mathcal{Q}_p$, $\mathcal{W}_c$. Let $R_{qc}, R_{qp}, R_{wc}, R_{wp}, R_1, R_2$ be positive rates s.t.:
\begin{align*}
    R_1 < (R + \epsilon) I(Q_p;Z|Q_c), \\
    R_2 < (R + \epsilon) I(X;Z|W_c).
\end{align*}
From $\mathcal{T}^n_{\delta}(Q_c)$ we pick $2^{kR_{qc}}$ sequences named $q^n_c(r_{qc})$, where $r_{qc} \in [1,2,...,2^{kR_{qc}}]$ is a index of a sequence.

For each $q^n_c(r_{qc})$ we pick $2^{k(R_{qp} + R_1)}$ sequences $q^n_p(r_{qc}, r_{qp}, r_1) \in \mathcal{T}^n_{\delta}(Q_p | q^n_c(r_{qc}))$.

Also for each $q^n_c(r_{qc})$ we randomly pick $2^{kR_{wc}}$ sequences $w^n_c(r_{qc}, r_{wc}) \in \mathcal{T}^n_{\delta}(W_c | q^n_c(r_{qc}))$.

And, finally, for each $q^n_c(r_{qc}), q^n_p(r_{qc}, r_{qp}, r_1), w^n_c(r_{qc}, r_{wc})$ we pick $2^{k(R_{wp} + R_2)}$ sequences $x^n (r_{qc},r_{qp},r_1,r_{wc},r_{wp},r_2) \in \mathcal{T}^n_{\delta}(X | q^n_c(r_{qc}), q^n_p(r_{qc}, r_{qp}, r_1), w^n_c(r_{qc}, r_{wc}))$.

This codebook is revealed to Bob and Eve.
Further, to shorten notations, we will skip indices in sequence names.

\textbf{Source encoding}. We have $(s^k, u^k)$ as encoder input. We search for the first jointly typical sequence $a^k_c$ s.t. $(a^k_c, s^k) \in \mathcal{T}^k_{\delta}(A_c,S)$.

Then, given $a^k_c$, we find first $a^k_p$ sequence s.t. $(a^k_p, s^k) \in \mathcal{T}^k_{\delta}(A_p,S | a^k_c)$.

Also, given codeword $a^k_c$, we proceed by finding first codeword $b^k_c$ s.t. $(b^k_c, u^k) \in \mathcal{T}^k_{\delta}(B_c, U | S, a^k_c)$.

And, given all previous codewords $a^k_c$, $a^k_p$ and $b^k_c$, we finish source encoding by finding first $(b^k_p, u^k) \in \mathcal{T}^k_{\delta}(B_p, U | S, a^k_c, a^k_p, b^k_c)$.

\textbf{Channel encoding}. We choose an arbitrary one-to-one mapping $(r_{qc},r_{qp},r_{wc},r_{wp}) = g(s_{ac},s_{ap},s_{bc},s_{bp})$ which is used to map source indices to channel indices. We assume that there exist mappings $(r_{qc},r_{qp}) = g_1(s_{ac}, s_{ap})$ and $(r_{wc},r_{wp}) = g_2(s_{bc}, s_{bp})$.

Now, given channel indexes $(r_{qc},r_{qp},r_{wc},r_{wp})$ as a result of mapping $g$, we sequentially select $q^n_c, q^n_p, w^n_c, x^n$, from the channel codebook. Alice transmit sequence $x^n \doteq x^n (r_{qc},r_{qp},r_1,r_{wc},r_{wp},r_2)$, where $r_1$ and $r_2$ selected at random with uniform distribution.

\textbf{Decoding}. Bob receives $y^n$. He sequentially searches in his codebook for codewords s.t.:
\begin{enumerate}
    \item $(q^n_c, y^n) \in \mathcal{T}^n_{\delta}(Q_c, Y)$,
    \item $(q^n_p, y^n) \in \mathcal{T}^n_{\delta}(Q_p, Y | q^n_c)$,
    \item $(w^n_c, y^n) \in \mathcal{T}^n_{\delta}(W_c, Y | q^n_c)$,
    \item $(x^n, y^n) \in \mathcal{T}^n_{\delta}(X, Y | q^n_c, q^n_p, w^n_c)$.
\end{enumerate}
Then, given channel indexes $(r_{qc},r_{qp},r_{wc},r_{wp})$, Bob using inverse mapping $g^{-1}$ gets source decoder indexes $(s_{ac},s_{ap},s_{bc},s_{bp})$ and decodes:
\begin{align*}
    \hat{s}^k &= \tilde{S} (a^k_c, a^k_p), \\
    \hat{u}^k &= \tilde{U} (a^k_c, a^k_p, b^k_c, b^k_p),
\end{align*}
where $\tilde{S} : A^k_c \times A^k_p \rightarrow \hat{S}^k$ and $\tilde{U} : A^k_c \times A^k_p \times B^k_c \times B^k_p \rightarrow \hat{U}^k$ are functions.
    
\textbf{Errors at encoding and decoding}.
We consider the following events which correspond to errors at the encoding or decoding stages.

Encoder errors:
\begin{align*}
    &\Err_1 \doteq
    \{
    \not\exists a^k_c : (a^k_c, s^k)
    \in \mathcal{T}^k_{\delta}(A_c,S)
    \}, \\
    &\Err_2 \doteq
    \{
    \not\exists a^k_p : (a^k_p, s^k)
    \in \mathcal{T}^k_{\delta}(A_p,S | a^k_c)
    \}, \\
    &\Err_3 \doteq 
    \{
    \not\exists b^k_c : (b^k_c, u^k) 
    \in \mathcal{T}^k_{\delta}(B_c,U | S, a^k_c)
    \}, \\
    &\Err_4 \doteq 
    \{
    \not\exists b^k_p : (b^k_p, u^k) 
    \in \mathcal{T}^k_{\delta}(B_p,U | S, a^k_c, a^k_p, b^k_c)
    \}.
\end{align*}
Decoder errors:
\begin{align*}
    &\Err_5 \doteq 
    \{
    \exists \hat{q}^n_c \neq q^n_c :
    \hat{q}_c^n, q^n_c \in \mathcal{T}^n_{\delta}(Q_c,Y)
    \}, \\
    &\Err_6 \doteq 
    \{
    \exists \hat{q}^n_p \neq q^n_p : \hat{q}^n_p, q^n_p  \in \mathcal{T}^n_{\delta}(Q_p,Y | q^n_c) 
    \}, \\
    &\Err_7 \doteq 
    \{
    \exists \hat{w}^n_c \neq w^n_c : \hat{w}^n_c, w^n_c  \in \mathcal{T}^n_{\delta}(W_c,Y | q^n_c) 
    \}, \\
    &\Err_8 \doteq 
    \{
    \exists \hat{x}^n \neq x^n : \hat{x}^n, x^n  \in \mathcal{T}^n_{\delta}(X, Y | q^n_c, q^n_p, w^n_c) \}. \\
\end{align*}

We upper-bound probability of an ``error" event as follows:
\begin{equation*}
    Pr \{\Err\} = P_{\Err} \leq  \sum_{i=1}^{8} P_{\Err_i},
\end{equation*}
where $P_{\Err_i} = Pr\{\Err_i\}$.

Given $k \rightarrow \infty$, it can be shown that: 
\begin{enumerate}
    \item if $R_{ac} > I(A_c; S)$ \\ then
    $P_{\Err_1} \rightarrow 0$,
    \item if $R_{ap} > I(A_p; S | A_c)$ \\ then
    $P_{\Err_2} \rightarrow 0$,
    \item if $R_{bc} > I(B_c; U | S, A_c)$ \\ then
    $P_{\Err_3} \rightarrow 0$,
    \item if $R_{bp} > I(B_p; U | S, A_c,A_p,B_c)$ \\ then
    $P_{\Err_4} \rightarrow 0$,
    \item if $R_{qc} < (R + \epsilon) I(Q_c; Y)$ \\ then
    $P_{\Err_5} \rightarrow 0$,
    \item if $R_{qp} + R_1 < (R + \epsilon) I(Q_p; Y | Q_c)$ \\ then
    $P_{\Err_6} \rightarrow 0$,
    \item if $R_{wc} < (R + \epsilon) I(W_c; Y | Q_c)$ \\ then
    $P_{\Err_7} \rightarrow 0$,
    \item if $R_{wp} + R_2 < (R + \epsilon) I(X; Y | Q_c, Q_p, W_c)$ \\ then
    $P_{\Err_8} \rightarrow 0$.
\end{enumerate}

\textbf{Analysis of expected distortion for semantics:}

\begin{align*}
    &\EX d_s \left( S^k, \hat{S}^k \right)
    = \EX d_s \left( S^k, \hat{f}_s(Y^n) \right) \\
    &=^{(a)} P_{\Err} \EX 
    \left\{ d_s \left( S^k, \hat{f}_s(Y^n) \right) | \Err \right\} \\
    &\qquad + P_{\bar{\Err}} \EX \left\{ d_s \left( S^k, \hat{f}_s(Y^n) \right) | \bar{\Err} \right\} \\
    &\leq P_{\Err} d_{S,m} + P_{\bar{\Err}} \EX \left\{ d_s \left( S^k, \hat{f}_s(Y^n) \right) | \bar{\Err} \right\} \\
    &=^{(b)} P_{\Err} d_{S,m} + P_{\bar{\Err}} \EX \left\{
    d_s \left( S^k, \tilde{S} (A^k_c,A^k_p) \right) | \bar{\Err} \right\} \\
    &\leq^{(c)} P_{\Err} d_{S,m} + P_{\bar{\Err}} (1 + \epsilon_1) \EX \left\{
    d_s \left( S, \tilde{S} (A_c,A_p) \right) | \bar{\Err} \right\} \\
    &\leq^{(d)} P_{\Err} d_{S,m} + (1 + \epsilon_1) \EX
    d_s \left( S, \tilde{S} (A_c,A_p) \right),
\end{align*}
where $d_{S,m} = \max_{S^k,Y^n} d_s \left( S^k, \hat{f}_s(Y^n) \right)$, (a) due to law of total expectation, (b) because $\hat{f}_s(Y^n) = \tilde{S} (A^k_c,A^k_p)$ given no error occurs $\bar{\Err}$, (c) due to typical average lemma and $(s^k,a^k_c,a^k_p) \in \mathcal{T}^k_{\delta}(S,A_c,A_p)$, (d) due to $\EX(X|A) \leq \frac{\EX X}{P_A}$.

We conclude that the following is sufficient to satisfy distortion condition (\ref{eq:cond_for_sem_distortion}) for semantics:
\begin{equation*}
    D_s \geq \EX d_s \left( S, \tilde{S} (A_c,A_p) \right),
\end{equation*}
given $k \rightarrow \infty$

\textbf{Analysis of expected distortion for observation:}

\begin{align*}
    &\EX d_u \left( U^k, \hat{U}^k \right)
    = \EX d_u \left( U^k, \hat{f}_u(Y^n) \right) \\
    &=^{(a)} P_{\Err} \EX
    \left\{ d_u \left( S^k, \hat{f}_s(Y^n) \right) | \Err \right\} \\
    &\qquad + P_{\bar{\Err}} \EX \left\{ d_u \left( U^k, \hat{f}_u(Y^n) \right) | \bar{\Err} \right\} \\
    &\leq P_{\Err} d_{U,m} + P_{\bar{\Err}} \EX \left\{ d_u \left( U^k, \hat{f}_u(Y^n) \right) | \bar{\Err} \right\} \\
    &=^{(b)} P_{\Err} d_{U,m} + P_{\bar{\Err}} \EX \left\{
    d_u \left( U^k, \tilde{U} (A^k_c,A^k_p,B^k_c,B^k_p) \right) | \bar{\Err} \right\} \\
    &\leq^{(c)} P_{\bar{\Err}} (1 + \epsilon_1) \EX \left\{
    d_u \left( S, \tilde{U} (A^k_c,A^k_p,B^k_c,B^k_p) \right) | \bar{\Err} \right\} \\
    &\qquad + P_{\Err} d_{U,m}\\
    &\leq^{(d)} P_{\Err} d_{U,m} + (1 + \epsilon_1) \EX
    d_u \left( U, \tilde{U} (A_c,A_p,B_c,B_p) \right),
\end{align*}
where $d_{U,m} = \max_{U^k,Y^n} d_u \left( U^k, \hat{f}_u(Y^n) \right)$, (a) due to law of total expectation, (b) because $\hat{f}_u(Y^n) = \tilde{U} (A^k_c,A^k_p,B^k_c,B^k_p)$ given $\bar{\Err}$, (c) due to typical average lemma and $(u^k,a^k_c,a^k_p,b^k_c,b^k_p) \in \mathcal{T}^k_{\delta}(U,A_c,A_p,B_c,B_p)$, (d) due to $\EX(X|A) \leq \frac{\EX X}{P_A}$.

Thus, to satisfy distortion condition (\ref{eq:cond_for_src_distortion}) it is sufficient to have:
\begin{equation*}
    D_u \geq \EX d_u \left( U, \tilde{U} (A_c,A_p,B_c,B_p) \right),
\end{equation*}
given $k \rightarrow \infty$

\textbf{Equivocation analysis for observation}. Here we treat sequence indices $s_{**}$ and $r_{**}$ as random variables. We start by analysing equivocation for observation $U^k$: 
\begin{align} \nonumber
    &H(U^k | Z^n) = H(U^k|M_u,Z^n) + I(U^k;M_u|Z^n) \\ \nonumber
    &= H(U^k| M_u) + H(M_u|Z^n) - H(M_u|U^k,Z^n) \\
    &\quad - I(U^k;Z^n|M_u),
    \label{eq:eqv_src_terms}
\end{align}
where $M_u \doteq (s_{bc},s_{bp})$ is an encoded (by source encoder) message for observation.

First term of (\ref{eq:eqv_src_terms}):
\begin{align} \nonumber
    &H(U^k| M_u) = H(U^k| s_{bc},s_{bp})
    = H(S^k,U^k) \\ \nonumber
    &\quad - I(s_{bc},s_{bp};S^k,U^k) - H(S^k|U^k,s_{bc},s_{bp}) \\ \nonumber
    &=^{(a)} H(S^k,U^k) - H(s_{bc},s_{bp}) - H(S^k|U^k,s_{bc},s_{bp}) \\ \nonumber
    &= H(S^k,U^k) - H(s_{bc}) - H(s_{bp}) + I(s_{bc};s_{bp}) \\
    &\quad - H(S^k|U^k,s_{bc},s_{bp}),
    \label{eq:eqv_src_1st_term}
\end{align}
where (a) because $(s_{bc},s_{bp})$ is a function of $(s^k,u^k)$

For second term (\ref{eq:eqv_src_terms}) we have:
\begin{align} \nonumber
    &H(M_u|Z^n) = H(s_{bc},s_{bp}|Z^n)
    =^{(a)} H(r_{wc},r_{wp}|Z^n) \\ \nonumber
    &\geq^{(b)} H(X^n|r_{wc}) + H(Z^n|X^n) -H(X^n|r_{wc},r_{wp},Z^n) \\ \nonumber
    &\quad - H(Z^n|r_{wc}) \\ \nonumber
    &= H(X^n|r_{wc}) + H(Z^n|X^n)
    -H(X^n|r_{wc},r_{wp},Z^n) \\
    &\quad - H(Z^n|r_{qc},r_{wc}) - I(Z^n;r_{qc}|r_{wc}),
    \label{eq:eqv_src_2nd_term}
\end{align}
where (a) due to $g_2$ mapping, (b) for same reasons as in \cite[eq. 2.38]{liang2009}.
Substituting (\ref{eq:eqv_src_1st_term}) and (\ref{eq:eqv_src_2nd_term}) into (\ref{eq:eqv_src_terms}) we have:
\begin{align} \nonumber
    &H(U^k|Z^n) \geq
    H(S^k,U^k) - H(s_{bc}) - H(s_{bp}) + I(s_{bc};s_{bp}) \\ \nonumber
    &\quad - H(S^k|U^k,s_{bc},s_{bp}) + H(X^n|r_{wc}) + H(Z^n|X^n) \\ \nonumber
    &\quad - H(X^n|r_{wc},r_{wp},Z^n) - H(Z^n|r_{qc},r_{wc}) - I(Z^n;r_{qc}|r_{wc}) \\
    &\quad - H(s_{bc},s_{bp}|U^k,Z^n) - I(U^k;Z^n|s_{bc},s_{bp}).
    \label{eq:eqv_src_inter}
\end{align}
Now we reduce some terms in (\ref{eq:eqv_src_inter}).
\begin{align} \nonumber
    &H(S^k,U^k) - H(S^k|U^k,s_{bc},s_{bp}) - H(s_{bc},s_{bp}|U^k,Z^n) \\
    &=^{(a)} H(U^k) + I(S^k;s_{bc},s_{bp};Z^n|U^k),
    \label{eq:reduced1_src}
\end{align}
where (a) because $(s_{bc},s_{bp})$ is a function of $(s^k,u^k)$.
\begin{align} \nonumber
    &H(X^n|r_{wc},r_{wp},Z^n) + I(U^k;Z^n|s_{bc},s_{bp}) \\ \nonumber
    &\leq^{(a)} H(X^n|r_{wc},r_{wp},Z^n) + I(X^n;Z^n|r_{wc},r_{wp}) \\
    &= H(X^n|r_{wc},r_{wp}),
    \label{eq:reduced2_src}
\end{align}
where (a) due to data processing inequality and $g_2$ mapping.
\begin{align}
    I(Z^n;r_{qc}|r_{wc}) \leq H(r_{qc}|r_{wc}) = H(r_{qc}) - I(r_{qc};r_{wc}).
    \label{eq:reduced3_src}
\end{align}

Returning to (\ref{eq:eqv_src_inter}) with (\ref{eq:reduced1_src}), (\ref{eq:reduced2_src}) and (\ref{eq:reduced3_src}):
\begin{align} \nonumber
    &H(U^k|Z^n) \geq
    H(U^k) - H(s_{bc}) - H(s_{bp}) + I(s_{bc};s_{bp}) \\ \nonumber
    &\quad + H(X^n|r_{wc}) + H(Z^n|X^n) - H(Z^n|r_{qc},r_{wc}) \\ \nonumber
    &\quad - H(r_{qc}) + I(r_{qc};r_{wc}) - H(X^n|r_{wc},r_{wp}) \\
    &\quad + I(S^k;s_{bc},s_{bp};Z^n|U^k).
    \label{eq:eqv_src_pre_final}
\end{align}

Now we bound each term of (\ref{eq:eqv_src_pre_final}):
\begin{enumerate}
    \item $H(U^k) = k H(U)$ ($U^k$ is i.i.d.),
    \item $H(s_{bc}) \leq kR_{bc}$,
    \item $H(s_{bp}) \leq kR_{bp}$,
    \item $H(X^n|r_{wc}) \geq k(R_{qc} + R_{qp} + R_1 + R_{wp} + R_2) - 1 - \epsilon_1$
    
    ($X^n$ is nearly uniform and  \cite[Lemma 2.5]{liang2009}),
    \item $H(Z^n|X^n) = nH(Z|X)$
    
    (channel is memoryless),
    \item $H(Z^n|r_{qc},r_{wc}) \leq n H(Z|Q_c,W_c) + n\epsilon_2$
    
    (see \cite[eq. 2.50-2.52]{liang2009}),
    \item $H(r_{qc}) \leq kR_{qc}$,
    \item $H(X^n|r_{wc},r_{wp}) \leq k(R_{qc} + R_{qp} + R_1 + R_2)$,
    \item $I(s_{bc};s_{bp}) + I(r_{qc};r_{wc}) +  I(S^k;s_{bc},s_{bp};Z^n|U^k) \geq 0$.
\end{enumerate}
Collecting all terms we have:
\begin{align*}
    &H(U^k|Z^n) \geq
    k H(U) - kR_{bc} - kR_{bp} + kR_{qc} + kR_{qp} + kR_1 \\
    &\quad + kR_{wp} + kR_2 - 1 - \epsilon_1 + nH(Z|X) - n H(Z|Q_c,W_c) \\
    &\quad - n\epsilon_2
    - kR_{qc} - k(R_{qc} + R_{qp} + R_1 + R_2) \\
    &= k H(U) - kR_{bc} - kR_{bp} + kR_{wp} - kR_{qc} + nH(Z|X) \\
    &\quad - n H(Z|Q_c,W_c) - n\epsilon_3 - 1.
\end{align*}
Thus, to satisfy equivocation for source, it is sufficient to have:
\begin{align*}
    &\Delta_u \leq H(U) - R_{bc} - R_{bp} + R_{wp} - R_{qc} \\
    &\quad + (R+\epsilon) (H(Z|X) - H(Z|Q_c,W_c)) - \epsilon,
\end{align*}
given $k \rightarrow \infty$.

\textbf{Equivocation analysis for semantics}.
For semantic equivocation, we have:
\begin{align} \nonumber
    &H(S^k|Z^n) = H(S^k|M_s,Z^n) + I(S^k;M_s|Z^n) \\ \nonumber 
    &= H(S^k|M_s) + H(M_s|Z^n) - H(M_s|S^k,Z^n) \\
    &\quad - I(S^k;Z^n|M_s),
    \label{eq:eqv_sem_terms}
\end{align}
where $M_s \doteq (s_{ac},s_{ap})$ is an encoded (by source encoder) message for semantics.
First term of (\ref{eq:eqv_sem_terms}):
\begin{align} \nonumber
    &H(S^k| M_s) = H(S^k| s_{ac},s_{ap}) \\ \nonumber
    &= H(S^k,U^k) - I(s_{ac},s_{ap};S^k,U^k) - H(U^k|S^k,s_{ac},s_{ap}) \\ \nonumber
    &=^{(a)} H(S^k,U^k) - H(s_{ac},s_{ap}) - H(U^k|S^k,s_{ac},s_{ap}) \\ \nonumber
    &= H(S^k,U^k) - H(s_{ac}) - H(s_{ap}) + I(s_{ac};s_{ap}) \\
    &\quad - H(U^k|S^k,s_{ac},s_{ap}),
    \label{eq:eqv_sem_1st_term}
\end{align}
where (a) because $(s_{ac},s_{ap})$ is a function of $(s^k,u^k)$.
Second term of (\ref{eq:eqv_sem_terms}):
\begin{align} \nonumber
    &H(M_s|Z^n) = H(s_{ac},s_{ap}|Z^n)
    =^{(a)} H(r_{qc},r_{qp}|Z^n) \\ \nonumber
    &\geq^{(b)}  H(X^n|r_{qc}) + H(Z^n|X^n) \\
    &\quad - H(X^n|r_{qc},r_{qp},Z^n) - H(Z^n|r_{qc}),
    \label{eq:eqv_sem_2nd_term}
\end{align}
where (a) due to $g_1$ mapping, (b) see \cite[eq. 2.38]{liang2009}.
Gathering all terms, we have:
\begin{align} \nonumber
    &H(S^k|Z^n) \geq H(S^k,U^k) - H(s_{ac}) - H(s_{ap}) + I(s_{ac};s_{ap}) \\ \nonumber
    &\quad - H(U^k|S^k,s_{ac},s_{ap}) + H(X^n|r_{qc}) + H(Z^n|X^n) \\ \nonumber
    &\quad - H(X^n|r_{qc},r_{qp},Z^n) - H(Z^n|r_{qc}) \\
    &\quad - H(s_{ac},s_{ap}|S^k,Z^n)- I(S^k;Z^n|s_{ac},s_{ap}).
    \label{eq:eqv_sem_inter}
\end{align}

We reduce some terms in (\ref{eq:eqv_sem_inter}).
\begin{align} \nonumber
    &H(S^k,U^k) - H(U^k|S^k,s_{ac},s_{ap}) - H(s_{ac},s_{ap}|S^k,Z^n) \\
    &=^{(a)} H(S^k) + I(U^k;s_{ac},s_{ap};Z^n|S^k),
    \label{eq:reduced1_sem}
\end{align}
where (a) because $(s_{ac},s_{ap})$ is a function of $(s^k,u^k)$.
\begin{align} \nonumber
    &H(X^n|r_{qc},r_{qp},Z^n) + I(S^k;Z^n|s_{ac},s_{ap}) \\ \nonumber
    &\leq^{(a)} H(X^n|r_{qc},r_{qp},Z^n) + I(X^n;Z^n|r_{qc},r_{qp}) \\
    &= H(X^n|r_{qc},r_{qp}),
    \label{eq:reduced2_sem}
\end{align}
where (a) due to data processing inequality and $g_1$ mapping.
Substituting (\ref{eq:reduced1_sem}) and (\ref{eq:reduced2_sem}) into (\ref{eq:eqv_sem_inter}) we obtain:
\begin{align} \nonumber
    &H(S^k|Z^n) \geq H(S^k) - H(s_{ac}) - H(s_{ap}) + I(s_{ac};s_{ap}) \\ \nonumber
    &\quad + H(X^n|r_{qc}) + H(Z^n|X^n) - H(Z^n|r_{qc}) \\
    &\quad - H(X^n|r_{qc},r_{qp}) + I(U^k;s_{ac},s_{ap};Z^n|S^k).
    \label{eq:eqv_sem_pre_final}
\end{align}

Now we bound each term of (\ref{eq:eqv_sem_pre_final}):
\begin{enumerate}
    \item $H(S^k) = k H(S)$ ($S^k$ is i.i.d.),
    \item $H(s_{ac}) \leq kR_{ac}$,
    \item $H(s_{ap}) \leq kR_{ap}$,
    \item $H(X^n|r_{qc}) \geq k(R_{wc} + R_{qp} + R_1 + R_{wp} + R_2) - 1 - \epsilon_1$
    
    ($X^n$ is nearly uniform and  \cite[Lemma 2.5]{liang2009}),
    \item $H(Z^n|X^n) = nH(Z|X)$
    
    (channel is memoryless),
    \item $H(Z^n|r_{qc}) \leq n H(Z|Q_c) + n\epsilon_2$
    
    (see \cite[eq. (2.50-2.52)]{liang2009}),
    \item $H(X^n|r_{qc},r_{qp}) \leq k(R_{wc}+R_{wp}+R_1+R_2)$,
    \item $I(U^k;s_{ac},s_{ap};Z^n|S^k) \geq 0$, 
    \item $I(s_{ac};s_{ap}) \geq 0$.
\end{enumerate}

Returning to (\ref{eq:eqv_sem_pre_final}) we have:
\begin{align*}
    &H(S^k|Z^n) \geq k H(S) - kR_{ac} - kR_{ap} \\
    &\quad + k(R_{wc} + R_{qp} + R_1 + R_{wp} + R_2) - 1 - \epsilon_1 \\
    &\quad + nH(Z|X) - n H(Z|Q_c) - n\epsilon_2  \\
    &\quad - k(R_{wc}+R_{wp}+R_1+R_2) \\
    &= k H(S) - kR_{ac} - kR_{ap} + kR_{qp} \\
    &\quad + nH(Z|X) - n H(Z|Q_c) - n\epsilon_3 - 1.
\end{align*}

Thus, to satisfy equivocation for semantics, it is sufficient to have:
\begin{align*}
    &\Delta_s \leq H(S) - R_{ac} - R_{ap} + R_{qp} \\
    &\quad + (R + \epsilon)(H(Z|X) - H(Z|Q_c)) - \epsilon,
\end{align*}
given $k \rightarrow \infty$.

\textbf{Joint equivocation analysis}:
\begin{align} \nonumber
    &H(S^k,U^k|Z^n) = H(S^k,U^k|M) + H(M|Z^n) \\ \nonumber
    &\quad - I(S^k,U^k;Z^n|M) \\
    &=^{(a)} H(S^k,U^k|M) + H(M|Z^n),
    \label{eq:joint_eqv_init_terms}
\end{align}
where $M = (s_{ac},s_{bc},s_{ap},s_{bp})$ and (a) due to $(S^k,U^k) \rightarrow M \rightarrow Z^n$.

First term of (\ref{eq:joint_eqv_init_terms}):
\begin{equation*}
    H(S^k,U^k|M) =^{(a)} H(S^k,U^k) - H(M),
\end{equation*}
where (a) because $M$ is a function of $(S^k,U^k)$.

Second term of (\ref{eq:joint_eqv_init_terms}):
\begin{align*}
    &H(M|Z^n) \geq^{(a)} H(X^n|r_c) + H(Z^n|X^n) - H(X^n|r_c,r_p,Z^n) \\
    &\quad - H(Z^n|r_c),
\end{align*}
where $r_c = (r_{qc},r_{wc})$, $r_p = (r_{qp},r_{wp})$, (a) due to $g$ mapping, the fact that $H(r_c,r_p|X^n) = 0$ and $X^n \rightarrow Y^n \rightarrow Z^n$.

Returning to (\ref{eq:joint_eqv_init_terms}) we have:
\begin{align}
    \nonumber
    &H(S^k,U^k|Z^n) \geq H(S^k,U^k) - H(M) + H(X^n|r_c) \\
    &\quad + H(Z^n|X^n) - H(X^n|r_c,r_p,Z^n) - H(Z^n|r_c).
    \label{eq:eqv_joint_pre_final}
\end{align}

Now we bound each term of (\ref{eq:eqv_joint_pre_final}):
\begin{enumerate}
    \item $H(S^k,U^k) = k H(S,U)$ ($S^k$ and $U^k$ are i.i.d),
    \item $H(M) \leq k(R_{ac} + R_{bc} + R_{ap} + R_{bp})$,
    \item $H(X^n|r_c) \geq k(R_{qp} + R_{wp} + R_1 + R_2) - 1 - \epsilon_1$

    ($X^n$ is nearly uniform, \cite[Lemma 2.5]{liang2009}),
    \item $H(Z^n|X^n) = nH(Z|X)$,
    \item $H(X^n|r_c,r_p,Z^n) \leq \epsilon_2$

    (due to Fano's inequality, see \cite[eq. (2.49)]{liang2009}),
    \item $H(Z^n|r_c) \leq n H(Z|Q_c,W_c) + n\epsilon_3$

    (see \cite[eq. (2.50-2.52)]{liang2009}).
\end{enumerate}

Finally, we obtain:
\begin{align*}
    &\frac{1}{k} H(S^k,U^k|Z^n) \geq H(S,U) - R_{ac} - R_{bc} - R_{ap} - R_{bp} \\
    &\quad + (R + \epsilon) \left( H(Z|X) - H(Z|Q_c,W_c)\right) + \epsilon \\
    &\quad + R_{qp} + R_{wp} + R_1 + R_2.
\end{align*}

Thus, to satisfy joint equivocation, it is sufficient to have:
\begin{align*}
    &\Delta_{su} \leq H(S,U) - R_{ac} - R_{bc} - R_{ap} - R_{bp} + R_{qp} + R_{wp} \\
    &\quad + R_1 + R_2 + (R + \epsilon) \left( H(Z|X) - H(Z|Q_c,W_c) \right) - \epsilon.
\end{align*}

\textbf{Summary of conditions:}
\begin{equation*}
    \begin{cases}
        R_{**}, R_1, R_2 > 0 \\
        R_{ac} + R_{ap} = R_{qc} + R_{qp} \\
        R_{bc} + R_{bp} = R_{wc} + R_{wp} \\
        R_{ac} < R_{qc} \\
        R_{bc} < R_{wc} \\
        R_{ac} > I(A_c;S) \\
        R_{ap} > I(A_p;S | A_c) \\
        R_{bc} > I(B_c;U | S,A_c) \\
        R_{bp} > I(B_p;U | S,A_c, A_p, B_c) \\
        R_{qc} < R I(Q_c;Y) \\
        R_{qp} + R_1 < R I(Q_p; Y | Q_c) \\
        R_{wc} < R I(W_c; Y | Q_c) \\
        R_{wp} + R_2 < R I(X; Y | Q_c, Q_p, W_c) \\
        R_1 < R I(Q_p;Z|Q_c) \\ 
        R_2 < R I(X;Z|W_c) \\
        D_s \geq \EX d_S(S, \tilde{S}(A_c, A_p)) \\
        D_u \geq \EX d_U(U, \tilde{U}(A_c, A_p, B_c, B_p)) \\
        \Delta_s \leq H(S) - R_{ac} - R_{ap} + R_{qp} \\
            \qquad + (R + \epsilon)(H(Z|X) - H(Z|Q_c)) - \epsilon \\
        \Delta_u \leq H(U) - R_{bc} - R_{bp} + R_{wp} - R_{qc} \\
            \qquad + (R+\epsilon) (H(Z|X) - H(Z|Q_c,W_c)) - \epsilon \\
        \Delta_{su} \leq H(S,U) - R_{ac} - R_{bc} - R_{ap} - R_{bp} + R_{qp} + R_{wp} \\
        \qquad + R_1 + R_2 + (R + \epsilon) \left( H(Z|X) - H(Z|Q_c,W_c) \right) - \epsilon.
    \end{cases}
\end{equation*}
The above system of inequalities can be reduced to the following system:
\begin{equation*}
\begin{cases}
        I(S; A_c) < R I(Q_c;Y), \\
        I(S; A_c, A_p) < R I(Y; Q_c, Q_p), \\
        I(U; B_c | S,A_c) < R I(W_c;Y | Q_c), \\
        I(U; B_c | S,A_c) + I(U; B_p | S, A_c, A_p, B_c) < \\
        \qquad < R [ I(W_c; Y | Q_c) + I(X;Y | Q_c, Q_p, W_c) ] \\
        D_s \geq \EX d_S(S, \tilde{S}(A_c, A_p)), \\
        D_u \geq \EX d_U(U, \tilde{U}(A_c, A_p,B_c, B_p)), \\
        \Delta_s \leq H(S|Ac,Ap) + R( H(Z|X)-H(Z|Qc) ) \\
        \qquad + R I(Qp;Y|Qc), \\
\end{cases}
\end{equation*}
\begin{equation*}
\begin{cases}
        \Delta_u \leq H(U) - I(A_c,A_p;S)-I(B_c;U|A_c) \\ 
        \qquad - I(B_p;U|S,A_c,A_p,B_c) + R H(Z|X) \\ 
        \qquad - R H(Z|Q_c,W_c) + R I(Q_p;Y|Q_c) \\
        \qquad + RI(X;Y|Q_c,Q_p,W_c), \\
        \Delta_{su} \leq H(S,U) - I(S;A_c,A_p) - I(U;B_p|S,A_c,A_p,B_c) \\
        \qquad - I(U;B_c|S,A_c) + R (H(Z|X) - H(Z|Qc,Wc)) \\
        \qquad + R I(Q_p;Y|Q_c) + R I(X;Y|Q_c,Q_p,W_c).
    \end{cases}
\end{equation*}

The proof is complete.
    \end{appendices}

    \bibliographystyle{IEEEtran}  
    \bibliography{main}

\end{document}